\documentclass[12pt]{iopart}

\usepackage{epsfig}

\begin{document}

\title{Superconductivity in SnSb with natural superlattice structure}

\author{Bin Liu$^{1,2}$, Jifeng Wu$^{1,2}$, Yanwei Cui$^{2}$, Hangdong Wang$^{4}$, Yi Liu$^{3}$, Zhicheng Wang$^{3}$, Zhi Ren$^{2,1,3}$\footnote[1]{Electronic address: zhi.ren@wias.org.cn}, Guanghan Cao$^{3,5,6}$}

\address{$^{1}$Department of Physics, Fudan University, Shanghai 200433, P. R. China}
\address{$^{2}$Institute of Natural Sciences, Westlake Institute for Advanced Study, Westlake University, 18 Shilongshan Road, Hangzhou 310024, P. R. China}
\address{$^{3}$Department of Physics, Zhejiang University, Hangzhou 310027, P. R. China}
\address{$^{4}$Department of Physics, Hangzhou Normal University, Hangzhou 310036, P. R. China}
\address{$^{5}$State Key Lab of Silicon Materials, Zhejiang University, Hangzhou 310027, P. R. China}
\address{$^{6}$Collaborative Innovation Centre of Advanced Microstructures, Nanjing University, Nanjing 210093, P. R. China}

\date{\today}

\begin{abstract}

We report the results of electrical resistivity, magnetic and thermodynamic measurements on polycrystalline SnSb, whose structure consists of stacks of Sb bilayers and Sn$_{4}$Sb$_{3}$ septuple layers along the $c$-axis.
The material is found to be a weakly coupled, fully gapped, type-II superconductor with a bulk $T_{\rm c}$ of 1.50 K, while showing a zero resistivity transition at a significantly higher temperature of 2.48 K.
The Sommerfeld coefficient and upper critical field, obtained from specific heat measurements, are 2.29 mJ mol$^{-1}$ K$^{-2}$ and 520 Oe, respectively. Compositional inhomogeneity and strain effect at the grain boundaries are proposed as possible origins for the difference in resistive and bulk superconducting transitions. In addition, a comparison with the rock-salt structure SnAs superconductor is presented.
Our results provide the first clear evidence of bulk superconductivity in a natural superlattice derived from a topological semimetal.

\end{abstract}
\pacs{74.25.-q, 74.70.Ad, 74.78.Fk}
\maketitle

\section{\label{sec:level1}Introduction}
Recently, superconductors derived from topological materials have attracted a lot of attention because of their potential as topological superconductors that are useful in fault tolerant quantum computing \cite{TSCreview1,TSCreview2,Tscreview3}.
A common strategy to look for such kind of superconductors is to dope topological (crystalline) insulators \cite{Tscreview3}, which increases the carrier concentration and induces superconductivity (SC). Prominent examples include $A_{x}$Bi$_{2}$Se$_{3}$ ($A$ = Cu \cite{CuxBi2Se3discovery,CuxBi2Se3zerobias}, Sr \cite{SrxBi2Se3}, Nb \cite{NbxBi2Se3}), and Sn$_{1-x}$In$_{x}$Te \cite{Sn1-xInxTezerobias,Sn1-xInxTearpes}.
On the other hand, it has been shown that natural superlattice structure can be used to tailor the properties of topological materials by changing the constituent building blocks \cite{superlattice}.
For instance, Sb$_{2}$Te, which is composed of [Sb$_{2}$] bilayers and [Sb$_{2}$Te$_{3}$] quintuple layers, exhibits significantly different surface and bulk band structures from both the topological insulator Sb$_{2}$Te$_{3}$ and the topological semimetal Sb \cite{Sb2Te}.
In this context, it is of interest to see whether similar superlattices without intentional doping can support SC.

SnSb is one of the stable binary phases at room temperature in the Sn-Sb system \cite{Sn-Sbdiagram}, and has been studied intensively as an anode material for (Li/Na)-ion batteries over the past two decades \cite{battery1,battery2,battery3,battery4}.
Initially, the material was reported to form a rhombohedrally distorted rock-salt-like structure \cite{Sn-Sbdiagram}.
Until recently, an incommensurate modulation is found along the $c$-axis \cite{SnSbrevisited}, and the structure of SnSb can be viewed as intercalation of Sn layers between adjacent Sb layers, leading to stacking of [Sb$_{2}$] bilayers and [Sn-Sb-Sn-Sb-Sn-Sb-Sn] septuple-layers shown in Fig. 1 \cite{SnSbstructure}.
Notably, it was mentioned in a paper by Geller and Hull in 1960s that SnSb exhibits SC as in rock-salt structure SnAs, albeit with two superconducting transitions \cite{SCinSnSb}.
While SnAs has been confirmed to be a type-I superconductor experimentally \cite{SCinSnAs}, the nature of SC in SnSb remains unclear to date.

In this paper, we present a comprehensive study of the physical properties of SnSb.
It is found that the material is an $n$-type metal in the normal state, and undergoes a transition to zero resistivity at 2.48 K.
However, magnetic susceptibility and specific heat measurements indicate that bulk type-II SC is established at a considerably lower temperature of 1.50 K. Furthermore, the electronic specific heat jump of SnSb follows a weak coupling BCS-like behavior, pointing to a fully gapped superconducting state.
The normal-state and superconducting parameters are extracted and compared with those of SnAs. The origin of two superconducting transitions is also discussed.

\section{\label{sec:level1}Experimental}

Polycrystalline SnSb samples were synthesized by using a two-step method.
High-purity shots of Sn (99.99\%) and Sb (99.999\%) with the stoichiometric ratio of 1:1 were melted in sealed evacuated quartz tube at 900 $^{\rm o}$C for 48 h with intermittent shaking to ensure homogeneity, followed by quenching into cold water.
The resulting ingot was then annealed at 320 $^{\rm o}$C for another 48 h, and finally quenched into cold water.
The purity of the sample was checked by powder X-ray diffraction (XRD) using a PANalytical x-ray diffractometer with a monochromatic Cu-K$_{\alpha1}$ radiation at room temperature.
For consistency reason, all the physical property measurements were performed on samples obtained from the same ingot.
A part of the ingot was cut into regular-shaped samples for electrical resistivity, Hall coefficient and specific heat measurements, and the remaining part was crushed into powders.
The typical dimensions are 4 mm $\times$ 0.8 mm $\times$ 0.35 mm and 2 mm $\times$ 2 mm $\times$ 0.2 mm for transport (resistivity/Hall) and specific heat measurements, respectively.
The resistivity was measured by using a standard four-probe method.
Resistivity, Hall coefficient and specific heat measurements down to 1.8 K were carried out on a Quantum Design PPMS-9 Dynacool.
Specific heat measurements down to 0.5 K were preformed on a Quantum Design PPMS-9 Evercool II.
The two sets of specific heat data agree well within 5\% in the overlapped temperature range.
The dc magnetization down to 0.4 K was done on crushed powders with a commercial SQUID magnetometer (Quantum Design MPMS3).

\section{\label{sec:level1}Results and Discussion}

\subsection{\label{sec:level1}{Crystal Structure}}

Figure 2(a) shows the XRD pattern of the SnSb sample at room temperature, together with structure refinement profile using the JANA2006 program \cite{JANA2006}.
All the diffraction peaks can be well fitted with the (3 + 1)-dimensional superspace group $R$$\overline{3}$$m$(00$\gamma$) (No.166.1),
where $\gamma$ = 1.315 is the modulation $q$-vector component.
As exemplified in Fig. 2(b), first-order satellite reflections due to the incommensurate modulation are clearly visible
between the strong peaks of the $R$$\overline{3}$$m$ average structure.
The refined lattice parameters are $a$ = 4.331(2) {\AA} and $c$ = 5.352(2) {\AA} in the hexagonal setting, or $a$ = 3.072(1) {\AA} and $\alpha$ = 89.65$^{\circ}$ in the rhombohedral setting.
Note that the $\alpha$ angel is very close to 90$^{\circ}$, hence the structure of SnSb deviates only slightly from the cubic symmetry.
These results are in good agreement with previous reports \cite{SnSbrevisited,SnSbstructure,Sn-Sbdiagram}, and demonstrate high sample quality.
It is pointed out that the structure of Sb can also be described by the $R$$\overline{3}$$m$(00$\gamma$) space group with $\gamma$ = 1.5 \cite{SnSbstructure}.
Moreover, Sn is just before Sb in the periodic table, hence a strong spin-orbit coupling is expected in SnSb.
Taken together, it is tempting to speculate that SnSb and Sb have a similar topological character.

\subsection{\label{sec:level1}{Resistivity and Hall coefficient}}

The main panel of Fig. 3(a) shows the temperature dependence of resistivity ($\rho$) for the SnSb sample under zero field.
The $\rho$ value at room temperature is $\sim$28 $\mu$$\Omega$ cm, which lies in between that of pure Sn and pure Sb \cite{Snresistivity}.
With decreasing temperature, $\rho$ decreases linearly down to $\sim$50 K and then varies as $T^{2.4}$ at lower temperature,
which is typical of a normal metal.
Nevertheless, the low residual resistivity ratio ($\rho_{\rm 300 K}$/$\rho_{\rm 3 K}$) of $\sim$2.6 suggests the presence of significant disorder scattering, likely due to antisite occupation between Sn and Sb.
On further cooling below 2.8 K, $\rho$ drops rapidly to zero, evidencing a transition to the superconducting state [inset of Fig. 3(a)].
Here we define $T_{\rm c}^{\rho}$ = 2.48 K as the temperature corresponding to the midpoint of the resistive drop.
On the other hand, as shown in Fig. 3(b), the Hall resistivity depends linearly on the magnetic field, and the corresponding Hall coefficient is negative and nearly temperature independent.
These results are consistent with the metallic character of SnSb, and identify electrons as the dominant carriers.
Assuming a one-band model, we obtain an electron density of 1.9 $\times$ 10$^{22}$ cm$^{-3}$ at 1.8 K.

\subsection{\label{sec:level1}{Magnetic susceptibility and magnetization}}
Figure 4(a) shows the temperature dependence of the magnetic susceptibility $\chi$ measured on powder samples in both zero-field cooling (ZFC) and field cooling (FC) modes with an applied field of 4.8 Oe.
Here the demagnetization effect is taken into consideration assuming sphere-shaped grains with the demagnetization factor $N_{\rm d}$ = 1/3.
A diamagnetic signal is observed in both $\chi_{\rm ZFC}$ and $\chi_{\rm FC}$, and its onset temperature $\sim$2.9 K is very close to that of the resistive transition.
Below 2.5 K, a divergence is present between the ZFC and FC curve, indicating the presence of trapped flux.
On closer examination, it is noted that the diamagnetic transition is rather broad at the beginning, leading to a low shielding fraction of 9.3 \% at 1.7 K [inset of Fig. 4(a)].
With further decreasing temperature, however, both ZFC and FC data show a steep drop and become nearly flat below 1.3 K, which correspond to shielding and Meissner fractions of 106\% and 48\%, respectively.
Note that the linear interpolation of this strong diamagnetic signal intersects with that of the initial one at 1.66 K, which is one-third lower than $T_{\rm c}^{\rho}$.
This strongly suggest that the zero resistivity transition is not of bulk nature, as will be shown below.

Fig. 4(b) shows the field dependence of magnetization ($M$) of SnSb at 0.4 K. The data exhibits a small hysteresis loop, which corroborates that SnSb is a type-II superconductor with weak pinning.
As can be seen in the inset, the initial magnetization curve is linear, as expected for a Meissner state.
The lower critical field $B_{\rm c1}$ can be estimated from the deviation of the linearity, which gives the effective $B_{\rm c1}^{\ast}$(0) $\approx$ 8 Oe.
After corrected by the demagnetization factor, $B_{\rm c1}$(0) = $B_{\rm c1}^{\ast}$(0)/(1 $-$ $N_{\rm d}$) $\approx$ 12 Oe is obtained.

\subsection{\label{sec:level1}{Specific heat}}
Figure 5 summarizes the results from the specific heat ($c_{\rm p}$) measurements for SnSb.
As can be seen in Fig. 5(a), $c_{\rm p}$ varies smoothly across $T_{\rm c}^{\rho}$ while exhibits a sharp jump at $\sim$1.6 K.
With increasing field, the specific heat jump shifts towards lower temperature and becomes broadened, consistent with a superconducting transition.
At 500 Oe, the anomaly is almost suppressed, and the data can be analyzed by the Debye model $c_{\rm p}$/$T$ = $\gamma_{\rm n}$ + $\beta_{3}$$T^{2}$ + $\beta_{5}$$T^{4}$,
where $\gamma_{\rm n}$ and $\beta_{i}$ ($i$ = 3, 5) are the electronic and phonon specific-heat coefficients, respectively.
The best fit to the data below 3 K yields $\gamma_{\rm n}$ = 2.29 mJ mol$^{-1}$ K$^{-2}$, $\beta_{3}$ = 0.43 mJ mol$^{-1}$ K$^{-4}$, and $\beta_{5}$ = 0.009 mJ mol$^{-1}$ K$^{-6}$.
Once $\beta_{3}$ is known, the Debye temperature $\Theta_{\rm D}$ can be calculated using the equation $\Theta_{\rm D}$ = (12$\pi^{4}$$NR$/5$\beta_{3}$)$^{\frac{1}{3}}$, where $N$ = 2 and $R$ is the molar gas constant 8.314 J/mol K.
This gives $\Theta_{\rm D}$ = 208 K for SnSb.
On the other hand, as shown in the inset of Fig. 5(a), the $c_{\rm p}$ value exceeds the Dulong-Petit limit of 3$N$$R$ = 49.88 J/mol K at temperatures above 200 K,
suggesting that all the phonons are excited in this temperature range. Hence one may estimate $\Theta_{\rm D}$ $\sim$ 200 K, in good agreement with the above calculation.

Figure 5(b) shows the normalized electronic specific heat $C_{\rm el}$/$\gamma_{\rm n}$$T$ at zero field after subtraction of the phonon contribution.
The result further confirms the absence of a specific heat anomaly corresponding to the resistive transition.
More importantly, it turns out that the $C_{\rm el}$/$\gamma_{\rm n}$$T$ jump can be reproduced by the BCS theory \cite{BCS}, especially at low temperature region, and
the entropy balance determines the bulk superconducting transition temperature $T_{\rm c}^{\rm bulk}$ = 1.50 K.
This suggests that SnSb is fully gapped with a conventional $s$-wave pairing symmetry and a zero temperature gap value $\Delta$(0) = 0.22 meV.
Nevertheless, since the data is limited to $T$/$T_{\rm c}^{\rm bulk}$ $\geq$ 0.34, the possibility of multiband SC with a very small gap in one of the bands cannot be ruled out.
Therefore, $c_{\rm p}$ measurements at lower temperature would be necessary to draw a more concrete conclusion.

Assuming a phonon mediated pairing mechanism, the electron-phonon coupling constant, $\lambda_{\rm ph}$, can be calculated by the inverted McMillan formula \cite{Mcmillan},
\begin{equation}
\lambda_{\rm ph} = \frac{1.04 + \mu^{\ast} \rm ln(\Theta_{\rm D}/1.45\emph{T}_{\rm c})}{(1 - 0.62\mu^{\ast})\rm ln(\Theta_{\rm D}/1.45\emph{T}_{\rm c}) - 1.04},
\end{equation}
where $\mu^{\ast}$ is the Coulomb repulsion pseudopotential.
With an empirical value of $\mu^{\ast}$ = 0.13, $\lambda_{\rm ph}$ =0.52 and 0.58 are obtained for $T_{\rm c}^{\rho}$ and $T_{\rm c}^{\rm bulk}$, respectively, which implies that SnSb is a weakly coupled superconductor.
In addition, since $\gamma_{\rm n}$ is related to the bare density of states $N$(0) at the Fermi level through the relation $\gamma_{\rm n}$ = $\frac{1}{3}$$\pi$$^{2}$$k_{\rm B}$$^{2}$$N$(0)(1 + $\lambda_{\rm ph}$),
one can obtain $N$(0) = 0.64 states/eV per formula unit.

\subsection{\label{sec:level1}{Superconducting parameters}}
Figure 6 shows the magnetic field-temperature phase diagrams determined from resistivity and specific heat measurements. Here the resistive-transition criterion for $T_{\rm c}$ under different fields is the same as that at zero field (see the inset). Using the Werthamer-Helfand-Hobenberg (WHH) thoery \cite{WHH}, the upper critical field $B_{\rm c2}$ can be extrapolated to zero temperature, which gives $B_{\rm c2}^{\rho}$(0) = 1660 Oe and $B_{\rm c2}$(0) = 520 Oe for the two measurements, respectively. The Ginzburg-Landau (GL) coherence length $\xi_{\rm GL}$(0) is then calculated as $\xi_{\rm GL}$(0) = $\sqrt{\Phi_{0}/2\pi B_{\rm c2}(0)}$, where $\Phi_{0}$ = 2.07 $\times$ 10$^{-15}$ Wb is the flux quantum. This yields $\xi_{\rm GL}^{\rho}$(0) = 44.5 nm and  $\xi_{\rm GL}^{\rho}$(0) = 79.5 nm.
Also the equations $B_{\rm c1}$(0)/$B_{\rm c2}$(0) = (ln$\kappa_{\rm GL}$ + 0.5)/(2$\kappa_{\rm GL}^{2}$) \cite{kappa} and $B_{\rm c1}$(0) = ($\Phi_{0}$/4$\pi$$\lambda_{\rm eff}^{2}$)(ln$\kappa_{\rm GL}$ + 0.5) allow one to deduce the GL parameter $\kappa_{\rm GL}$ = 7.4 and the effective penetration depth $\lambda_{\rm eff}$ = 585 nm. Note that this $\lambda_{\rm eff}$ is more than two orders of magnitude smaller than the typical grain size of about 150$-$200 $\mu$m, reassuring that the effect of penetration depth is negligible \cite{SVF}.

\subsection{\label{sec:level1}{Origin of two superconducting transitions}}
From the above results, it is clear that there exists two superconducting transitions at $T_{\rm c}^{\rho}$ = 2.48 K and $T_{\rm c}^{\rm bulk}$ = 1.50 K in SnSb, respectively.
This is in agreement with the previous report \cite{SCinSnSb}, although the two $T_{c}$ values are slightly higher in the present case,
which is likely due to the different sample preparation conditions.
We emphasize that since no additional peaks are observed in the XRD data, the SC at $T_{\rm c}^{\rho}$ is unlikely due to a secondary phase with different structure.
Actually, this marked difference between $T_{\rm c}^{\rho}$ and $T_{\rm c}^{\rm bulk}$ is reminiscent of what has been found in CeIrIn$_{5}$ \cite{CeIrIn5}, PrOs$_{4}$Sb$_{12}$ \cite{PrOs4Sb12}, CePt$_{3}$Si \cite{CePt3Si}, La$_{3}$Rh$_{4}$Sn$_{13}$ \cite{La3Rh4Sn3}, and SrTi$_{1-x}$Nb$_{x}$O$_{3}$ \cite{SrTiNbO3}.
Note that the small $\gamma_{\rm n}$ value of SnSb is indicative of a weak electron correlation, which precludes the formation of a textured superconducting phase due to a competing order \cite{pressurizedCeRhIn5}.
Instead, two possibilities are considered here.

One possibility is that the existence of two superconducting transitions is due to the micro-phase segregation with a spread in the Sn/Sb ratio, which has been observed in long-term annealed SnSb samples \cite{SnSbrevisited}. These off-stoichiometric regions usually have a characteristic length of a few hundred $\mu$m or smaller and a very similar structure to SnSb \cite{Sn-Sbdiagram}, and thus are difficult to distinguish by the XRD measurements. Although a much shorter annealing time is adopted in the present study, it is possible that these inhomogeneous regions are already present, which is responsible for the resistive transition at $T_{\rm c}^{\rho}$, as in La$_{3}$Rh$_{4}$Sn$_{3}$ \cite{La3Rh4Sn3}. Moreover, the initial diamagnetic signal in the susceptibility data could also be due to these regions.
Hence we take the shielding fraction of $\sim$13\% at 1.66 K, the temperature at which the linear interpolation of the strong diamagnetic signal intersects with the base line [see Fig. 4(a)], as an estimation of the upper limit of their volume faction.
This indicates that such inhomogeneous regions, if they exist, occupy only a small part of the sample, in consistent with the absence of an anomaly in the specific heat data near $T_{\rm c}^{\rho}$.
Hence their presence has no practical effect in determining the intrinsic properties of SnSb.

Another possibility is that the zero resistivity transition results from the strain effect at the grain boundaries.
It is prudent to note that SnSb is not a line compound \cite{Sn-Sbdiagram}, and hence rapid quench is necessary to retain the phase to room temperature. During this process, it is expected that the grain surface cools faster than the interior, which leads to the development of strain at the grain boundaries.
This may result in the increase of density of states and/or phonon softening, and consequently an enhanced $T_{\rm c}$.
Note that the grain boundaries constitute a small fraction of the sample, yet they are connected to form a continuous superconducting path, as observed experimentally. In this respect, high pressure study and epitaxial film growth of SnSb, which may favor a higher $T_{\rm c}$ \cite{enhancedSCinSnAs,enhancedSCinSnSb}, will be of interest in future.

\subsection{\label{sec:level1}{Comparison with SnAs}}
Finally, we present a comparison of the normal-state and superconducting properties between SnSb and SnAs \cite{SCinSnAs}, which are summarized in Table 1.
Despite their different crystal structures, the $\gamma_{\rm n}$ and $\Theta_{\rm D}$ values are very similar for these compounds. Nevertheless, compared with SnAs, SnSb has a much lower $T_{\rm c}$ and a smaller $\lambda_{\rm ph}$.
Actually, this is also the case when one compares the experimental results of SnSb with the theoretically predicted ones assuming the same crystal structure as SnAs \cite{enhancedSCinSnSb}. This implies that, within the BCS framework, the lower $T_{\rm c}$ value of SnSb in the rombohedral structure compared with that in the rock-salt structure is due to a decrease in the electron phonon coupling strength.
However, it is noted that the electronic band structure of SnAs measured by ARPES is not in full agreement with the calculated one, but bears signatures of the spin-orbit coupling (SOC) \cite{SnAsarpes}.
As for SnSb, the SOC effect is expected to be even stronger, and hence the possibility of odd-parity pairing \cite{CuxBi2Se3zerobias,Sn1-xInxTezerobias} cannot be excluded.
In this regard, to better understand the SC in SnSb, a combined theoretical and spectroscopic study of its band structure and Fermi surface is strongly called for.

\section{\label{sec:level1}Conclusion}
In summary, we have studied the electrical transport, magnetic and thermodynamic properties of SnSb, which confirms the occurrence of SC in this natural superlattice phase. It is further shown that the material a bulk type-II superconductor below 1.50 K with a BCS-like gap. Nevertheless, the zero resistivity transition takes place at a considerably higher temperature of 2.48 K, for which two possibilities are discussed.
One possibility is that this is due to the presence of inhomogeneous regions with off stoichiometric Sn/Sb compositions due to the micro-phase segregation.
Another possible scenario in that the enhanced $T_{\rm c}$ in resistivity measurement results from the strain effect that develops at the grain boundaries during the sample quenching.
We also compare the properties of SnSb with those of the rock-salt structure SnAs superconductor, which suggests that the lower $T_{\rm c}$ in SnSb is likely due to a weakening of the electron phonon coupling strength.
Since SnSb can be regarded as a close relative to the topological semimetal Sb, further band structure calculations and surface-sensitive spectroscopic studies are called for to assess the potential of this material as a topological superconductor.

\section*{Acknowledgments}
We thank Xiao Lin for useful discussion. This work is supported by the National Key Research and Development Program of China (No.2017YFA0303002) and the Fundamental Research Funds for the Central Universities of China.

\pagebreak[4]

\begin{figure}
\centering
\includegraphics[width=12cm]{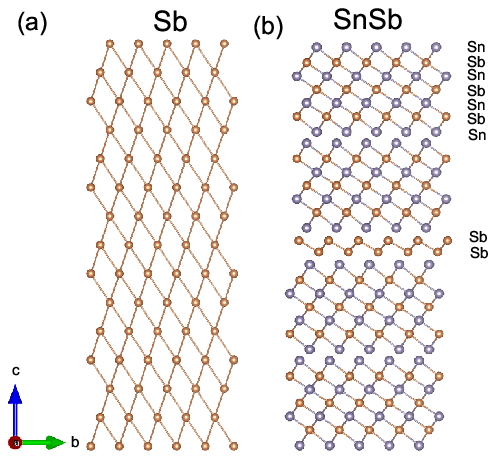}
\caption{\label{Fig.1} (color online). Schematic structure of (a) Sb and (b) SnSb projected along the $a$-axis following Ref. \cite{SnSbstructure}. }
\end{figure}

\begin{figure}[h]
\centering
\includegraphics[width=12cm]{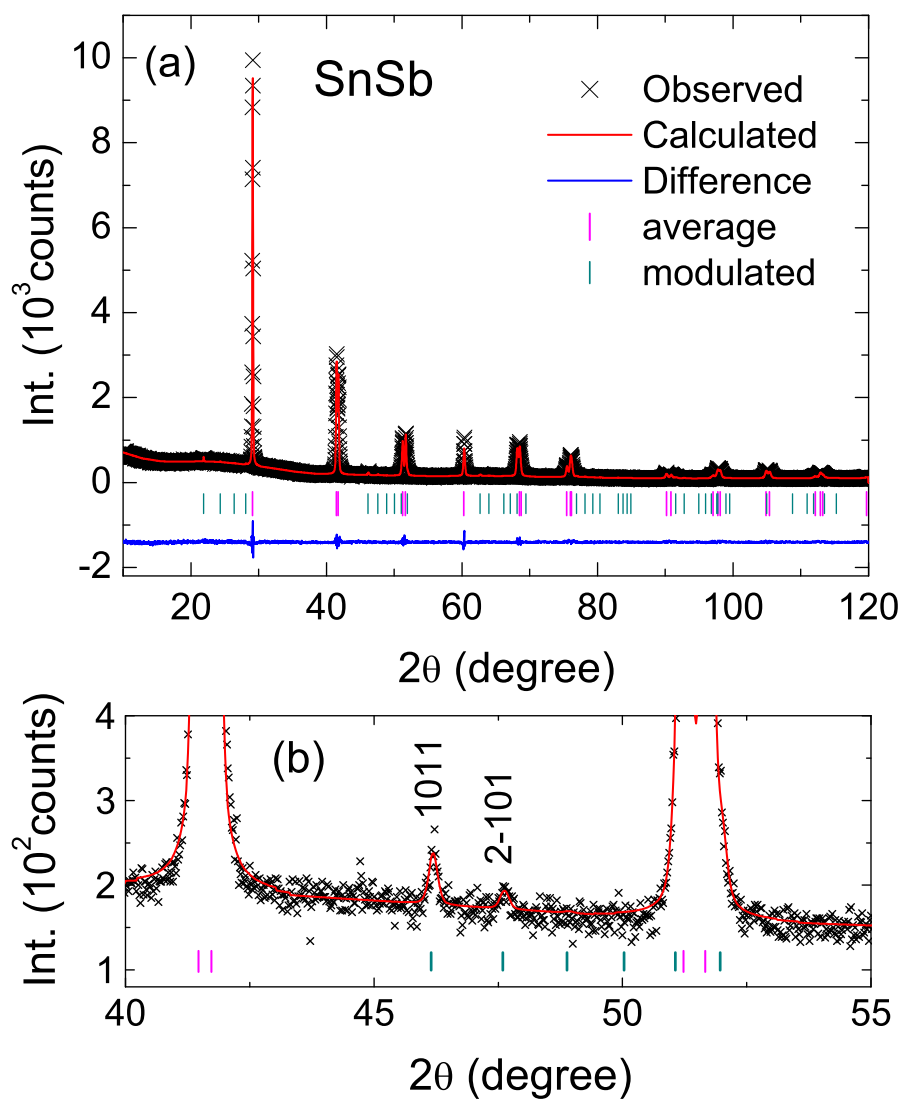}
\caption{\label{label} (a) Powder x-ray diffraction pattern and its refinement profile at room temperature for SnSb. Atomic position: Sn (0, 0, 0) and Sb (0, 0, 0). Refinement results: $R_{\rm wp}$ = 6.95 \%, $R_{\rm p}$ = 5.03 \%, goodness-of-fit GOF = 1.10.
(b) A zoom of the pattern between 40$^{\circ}$ and 53$^{\circ}$, showing the satellite diffraction peaks due to the incommensurate structural modulation.}
\end{figure}

\begin{figure}[h]
\centering
\includegraphics[width=12cm]{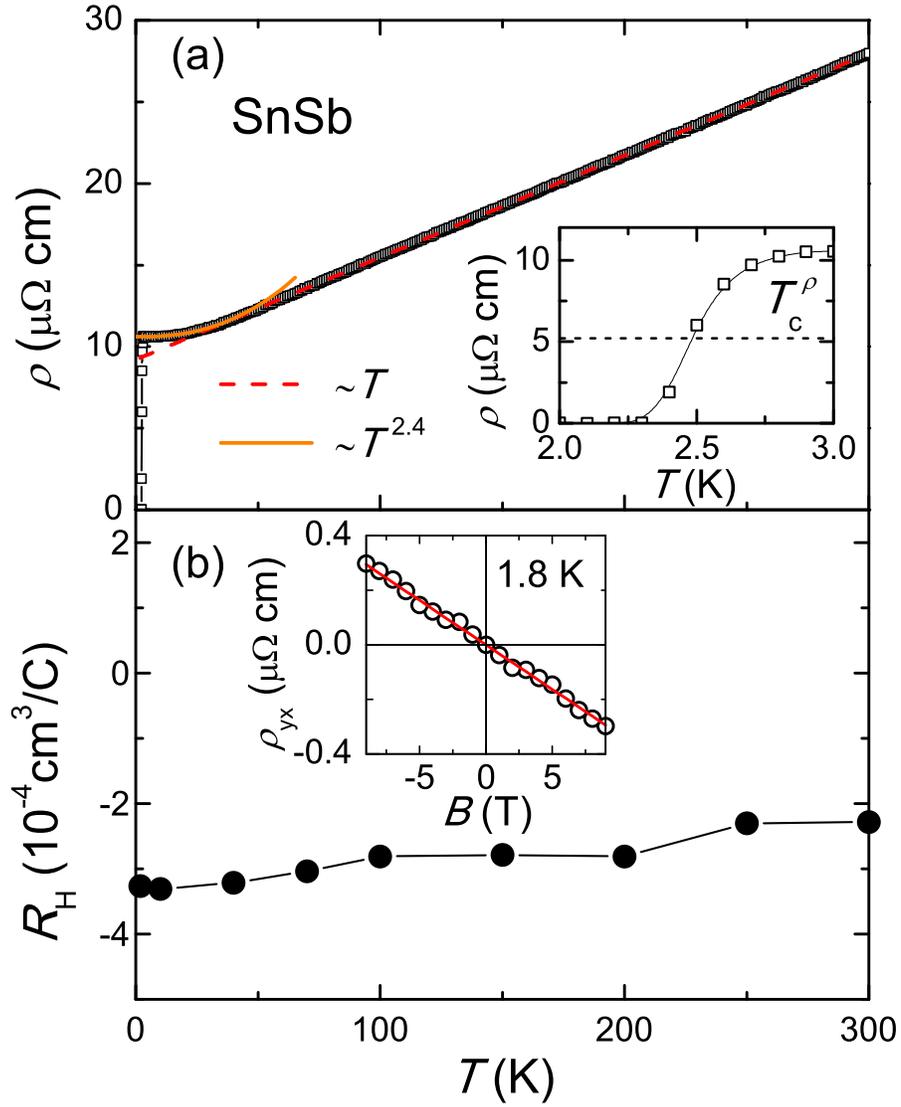}
\caption{\label{label}(a) Temperature dependence of resistivity for SnSb. The dashed and solid lines are the linear and power-law fits to the data of different temperature intervals.
The lower right inset shows an enlarged view of the low-temperature region. The dashed line marks the level corresponding to 50\% of the normal-state resistivity.
(b) Temperature dependence of Hall coefficient for SnSb. The inset shows the magnetic-field dependence of Hall resistivity at 1.8 K.
The solid line is a linear fit to the data.}
\end{figure}

\begin{figure}[h]
\centering
\includegraphics[width=12cm]{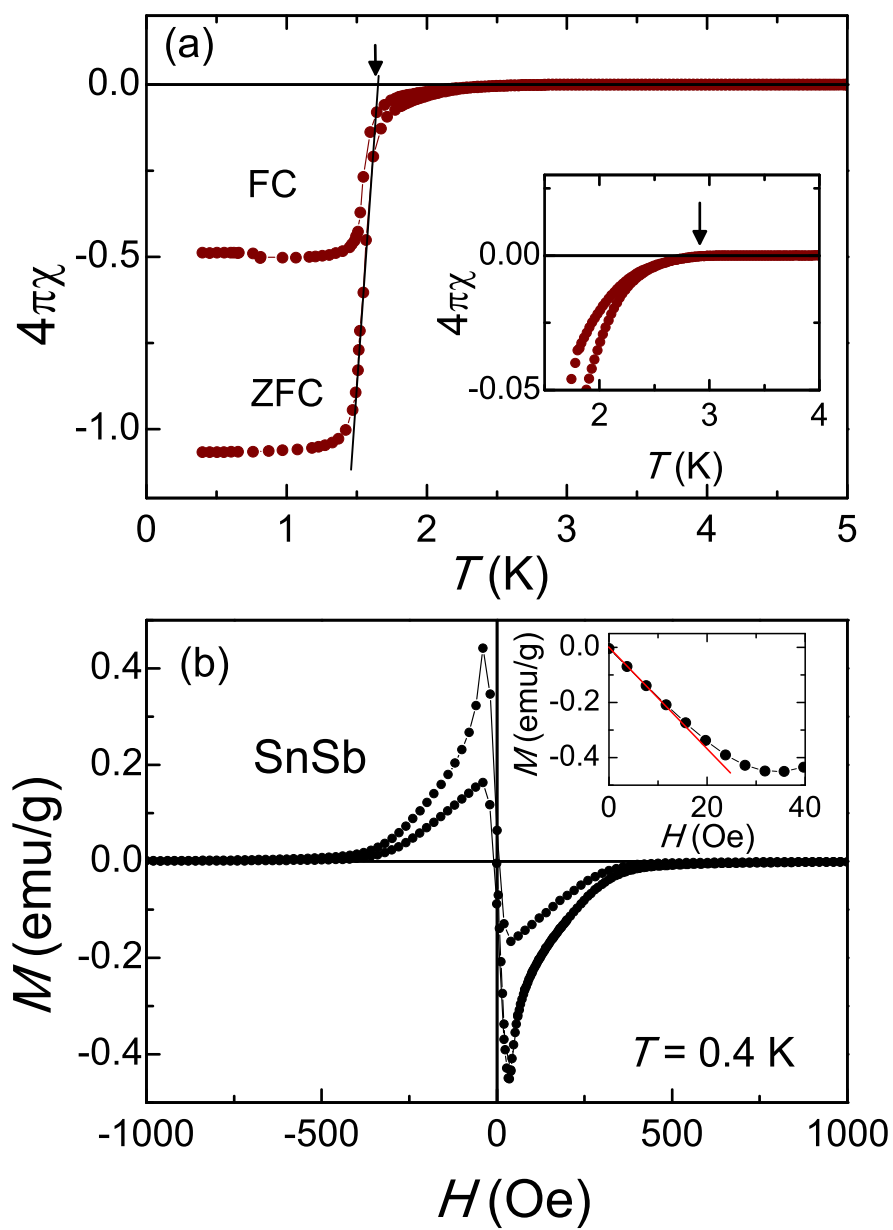}
\caption{\label{label} (a) Temperature dependence of magnetic susceptibility measured under $H$ = 4.8 Oe for SnSb.
The solid line and the arrow are a guide to the eyes.
The inset shows an enlarged view of the data near the onset of diamagnetic transition, which is indicated by the arrow.
The solid line denotes the baseline.
(b) Field dependence of magnetization at 0.4 K. The upper-right inset shows the initial magnetization curve.
The solid line is a guide to the eyes.}
\end{figure}

\begin{figure}[h]
\centering
\includegraphics[width=12cm]{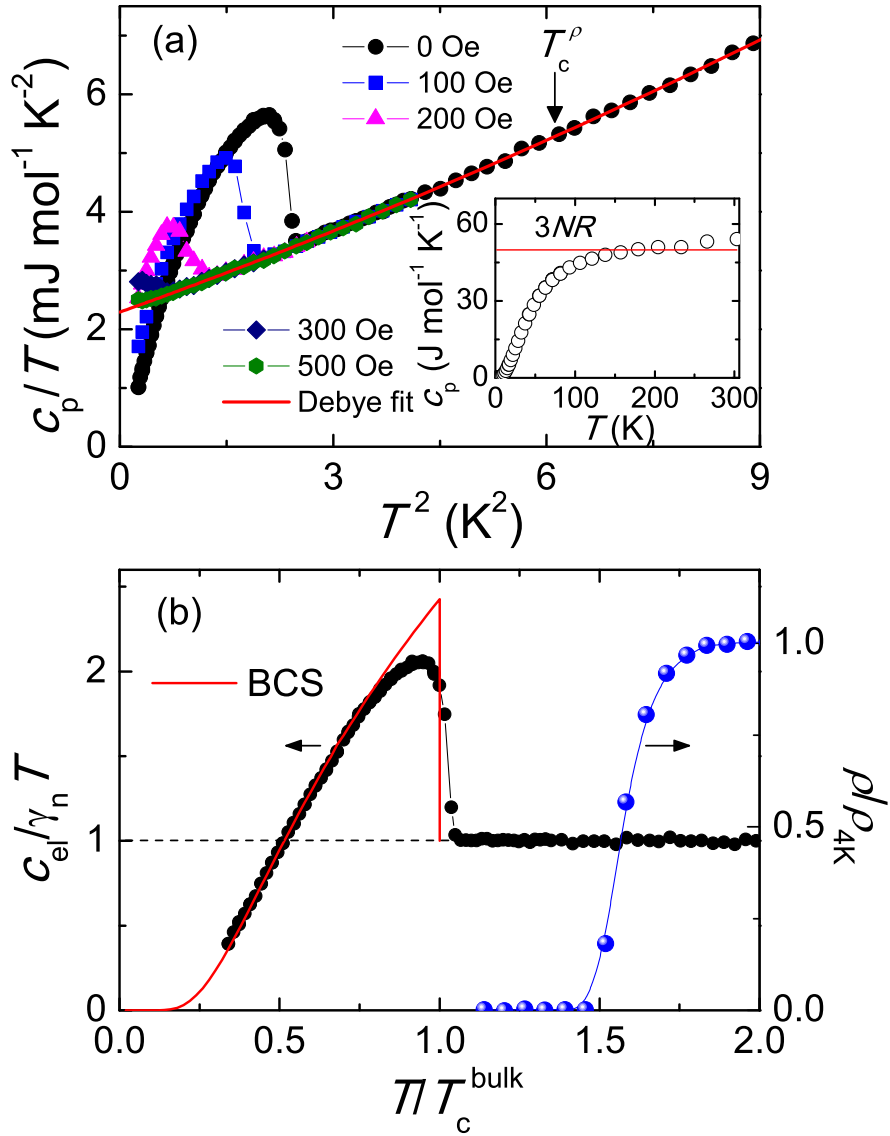}
\caption{\label{label} (a) Temperature dependence of specific heat for SnSb below 3 K at different fields plotted as $c_{\rm p}$/$T$ versus $T^{2}$.
The solid line is a fit by the Debye model to the data (see text), and $T_{\rm c}^{\rho}$ is marked by the arrows.
The inset shows the $c_{\rm p}$ data up to 300 K.
The solid line denotes the value of 3$N$$R$ = 49.88 J/molK.
(b) Normalized electronic specific heat at zero field.
The solid line is the curve given by the weak-coupling BCS theory, and the dashed line is a guide to the eyes.
For comparison, the resistivity transition data is also included (right axis). }
\end{figure}

\begin{figure}[h]
\centering
\includegraphics[width=12cm]{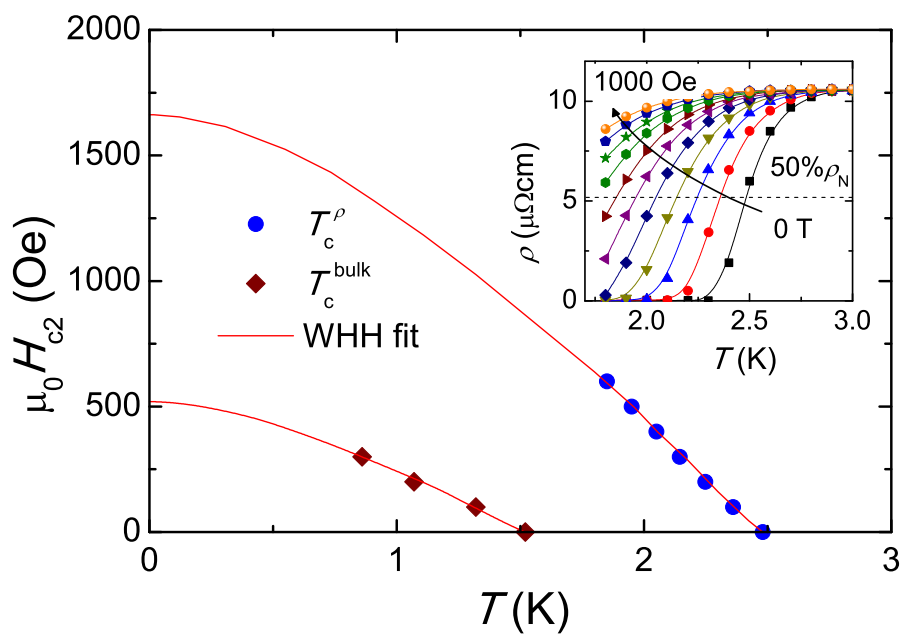}
\caption{\label{label} Temperature dependence of the critical fields determined from the resistivity and specific heat measurements.
The solid lines are WHH fits to the data.
The inset shows the temperature dependencies of resistivity under magnetic fields up to 1000 Oe in increments of 100 Oe.
The dashed line marks the level corresponding to 50\% of the normal-state resistivity.}
\end{figure}

\clearpage
\begin{table}
\caption{Comparison of the normal-state and bulk superconducting parameters between rombohedral SnSb and rock-salt SnAs \cite{SCinSnAs}.
Here $T_{\rm c}$ is the superconducting transition temperature, $B_{\rm c1}$(0) the zero-temperature lower critical field for type-II superconductors, $B_{\rm c2}$(0) the zero-temperature upper critical field for type-II superconductors,
$B_{\rm c}$(0) the zero-temperature critical field for type-I superconductors, $\lambda_{\rm eff}$ the effective penetration depth, $\xi_{\rm GL}$(0) the zero-temperature Ginzburg-Landau coherence length, $\kappa_{\rm GL}$ the Ginzburg-Landau parameter, $\Theta_{\rm D}$ the Debye temperature, $\gamma_{\rm n}$ the Sommerfeld coefficient, $\lambda_{\rm ph}$ the electron-phonon coupling constant, and $N$(0) the bare density of states at the Fermi level.
Note that the values of SnSb are mostly based on the specific heat measurements, except for $B_{\rm c1}$(0) that is deduced from the field dependence of magnetization at 0.4 K. }

\begin{indented}
\lineup
\item[]\begin{tabular}{@{}*{4}{l}}
\br
\0\0Parameters&SnSb &\0\0\0\0\0\0\0SnAs\\
\mr
\0\0$T_{\rm c}$ (K)							 &  1.50  & \0\0\0\0\0\0\03.58		\\
\0\0$B_{\rm c1}$(0)  (Oe)				& 12	& \0\0\0\0\0\0\0\0$-$ \\
\0\0$B_{\rm c2}$(0)  (Oe)				& 520	& \0\0\0\0\0\0\0\0$-$ \\
\0\0$B_{\rm c}$(0)  (Oe)				& \0$-$	& \0\0\0\0\0\0\0172 \\
\0\0$\lambda_{\rm eff}$  (nm)			 & 	  585	& \0\0\0\0\0\0\0\0$-$	\\
\0\0$\xi_{\rm GL}$(0)  (nm)				 & 	  79.5	& \0\0\0\0\0\0\0\0$-$	\\
\0\0$\kappa_{\rm GL}$			& 	 7.4	& \0\0\0\0\0\0\00.05	\\
\0\0$\Theta_{\rm D}$ (K)				 & 			208 	& \0\0\0\0\0\0\0235		 \\
\0\0$\gamma_{\rm n}$ (mJ mol$^{-1}$ K$^{-2}$)	&   2.29  & \0\0\0\0\0\0\02.18 \\
\0\0$\lambda_{\rm ph}$	 &   0.52 & \0\0\0\0\0\0\00.62  \\
\0\0$N$(0) (states/eV.f.u.)	& 0.64 & \0\0\0\0\0\0\0\0$-$  \\
\br
\end{tabular}
\end{indented}
\end{table}

\end{document}